\documentclass[aps,prl,twocolumn,superscriptaddress]{revtex4}
\usepackage{graphicx}
\def\resxx{\mbox{Re$(\sigma_{xx})$}}
\begin{document}


\title{Pinning modes and interlayer  
correlation in high magnetic field  bilayer Wigner solids}  



\author{Zhihai Wang}
\affiliation{National High Magnetic Field Laboratory, 1800 E. Paul Dirac Drive, Tallahassee, FL 32310}
\affiliation{Department of Physics, Princeton University, Princeton, NJ 08544}
\author{Yong P. Chen}
\altaffiliation[Current address: ]{Rice University, Houston, TX}
\affiliation{National High Magnetic Field Laboratory, 1800 E. Paul Dirac Drive, Tallahassee, FL 32310}
\affiliation{Department of Electrical Engineering, Princeton University, Princeton, NJ 08544}
\author{L. W. Engel}
\affiliation{National High Magnetic Field Laboratory, 1800 E. Paul Dirac Drive, Tallahassee, FL 32310}
\author{D. C. Tsui}
\affiliation{Department of Electrical Engineering, Princeton University, Princeton, NJ 08544}
\author{E. Tutuc}
\altaffiliation[Current address: ]{University of Texas, Austin, TX}
\affiliation{Department of Physics, Princeton University, Princeton, NJ 08544}
\author{M. Shayegan}
\affiliation{Department of Electrical Engineering, Princeton University, Princeton, NJ 08544}


\date{\today}

\begin{abstract}
We report  studies of  pinning mode resonances in the low total Landau filling ($\nu$) Wigner solid of a series of bilayer hole samples with negligible interlayer tunneling, and with varying interlayer separation $d$.  Comparison of states with equal layer densities $(p,p)$    to   single layer states $(p,0)$ produced {\em in situ} by biasing, indicates that there is    interlayer quantum correlation in the solid at small $d$.  Also, the resonance frequency at small $d$ is decreased just near $\nu=1/2$ and $2/3$, indicating the importance in the solid of  correlations related to those in the fractional quantum Hall effects. 
\end{abstract}

\pacs{73.43.Ðf,73.63.Hs}

\maketitle

Bilayers, closely spaced sheets of carriers in semiconductor hosts,  exhibit phenomena           that arise from interlayer interaction and    quantum correlation and so are not present in single layers. 
Of particular importance is the the quantum Hall  (QH) state at total bilayer Landau filling factor $\nu=1$, for the case of small interlayer separation and 
weak interlayer tunneling.   Owing to striking phenomena, both 
in magnetotransport \cite{eisnu1,tutucnu1} and in interlayer tunneling  
\cite{spielman},  that state is understood  to have carrier wave functions  which extend coherently between the two layers, with  the difference in the quantum  
 phase  between the layers spontaneously developing long range spatial coherence over the plane.    This $\nu=1$ state can be thought of as bilayer exciton condensate, excitonic counterflow superfluid,   or  a  pseudospin ferromagnet with  magnetization in the plane.  
As usual in describing bilayer states,   pseudospin specifies  the layer.    
   
Interlayer  quantum correlation, in the sense of wave functions spreading coherently between layers when tunneling is small, is  present in bilayers  at other $\nu$ than one.   Such interlayer correlation   underlies the    $\nu=1/2$ \cite{eis12,suen12,suen1232} and $3/2$ \cite{suen1232,hamilton32}   fractional QH effects (FQHE) observed in bilayers.  
This work will focus on  interlayer correlation  within  bilayer Wigner crystal (BWC) phases \cite{manoharan,tutucnu1imbal,faniel}, in samples with negligible interlayer tunneling.       Wigner crystals, lattices stabilized by  repulsion between carriers, have been  predicted in the absence of disorder \cite{ho,zhengfertig} for bilayers, as well as for single layer     systems \cite{wcpredict},  at the low $\nu$ termination of the  QH series.

  Theories \cite{ho,zhengfertig,peeters} have predicted a number of distinct BWC phases.    The relative  importance of interlayer  and intralayer interaction is crucial in these theories, and is measured by  the ratio  $d/a$,  where $d$ is the center-to-center layer separation,  $a=(2\pi p)^{-1/2}$ is the mean in-plane carrier spacing and $p$ is the carrier density per layer.     A one component triangular lattice   is expected at small enough $d/a$, and is an easy plane  pseudospin ferromagnetic   BWC (FMBWC) with one carrier, evenly and  coherently spread between the two layers, at each lattice site.    Interlayer-staggered two component lattices  occur  at larger  $d/a$, and   without interlayer tunneling, are pseudospin antiferromagnetic BWCs (AFMBWC), with carriers essentially completely in one layer alternating with those completely in the other.    Multiple staggered two component phases  are predicted,  including square and rectangular lattices. 
    For  $d\gg a$, the intralayer interaction dominates,  so the layers are simply  triangular lattices like single layer Wigner crystals, but interlayer-staggered.  
 
Wigner crystals in real samples, bilayer   or single layer, are pinned by disorder, and  so  are insulators.   Pinning also produces a striking microwave or rf conductivity resonance,  or  ``pinning mode", which is  a collective  oscillation of the  carriers about their pinned positions.     Pinning mode resonances of single layers   have been   studied  experimentally \cite{ssc06,yongab,yongmelt,yewc,clibdep,clidensity} and theoretically  
\cite{chitra,fertig,fogler}   and have proven to be   valuable  for obtaining information about single layer, pinned Wigner solids.   The  resonance frequency  always increases as the disorder strength increases,  and is 
also sensitive \cite{chitra,fertig,fogler} to the correlation length of  the effective disorder,   which takes into account the spread of     the carrier  wave functions.    Pinning modes of   bilayers \cite{doveston}  have also been observed.

We present a systematic study of the pinning mode in a  series of bilayer  samples, which have  negligible interlayer  tunneling   and  widely varying  layer separations.      For comparison with work on the $\nu=1$ excitonic condensate  \cite{eisnu1,tutucnu1,spielman},   we  present the effective  layer separation  as  $\tilde{d}=d/l_B=2^{1/2} d/a$,  where $l_B$ is  the magnetic length at $\nu=1$ in the balanced state.     
To isolate effects of  interlayer interaction or quantum correlation, we compare spectra  of  balanced states (layer carrier densities
 $(p,p)$) to those of  single layers (layer densities $(p,0)$)  created {\em in situ}  by  depleting one of the layers.  
Denoting  the resonance frequencies in these states by  $f_{pp}$ and $f_{p0}$,
  we focus on $\eta=f_{pp}/f_{p0}$ vs $\tilde{d}$, which has   
   a distinct minimum  at  $\tilde{d}\approx 1.8$.
    The interpretation  is in terms of two competing effects, which respectively tend to lower and raise $\eta$ as $\tilde{d}$  decreases: 
 1)      turn-on of the interlayer interaction on going from the $(p,0)$ single layer to the   $(p,p)$ BWC   and 2)   enhancement  \cite{yong} of the  effective pinning disorder in the  $(p,p)$ state   relative to that in the $(p,0)$  state {\em only when the $(p,p)$ state is an interlayer quantum correlated FMBWC}.  
 Of importance as well, and  present only for  small $\tilde{d}$,
are distinct dips in $f_{pp}$ vs $B$, around $\nu=1/2 $ or  $2/3$, demonstrating the effect, within  the low $\nu$ BWC insulator, of correlations  present in the FQHE states.

The  samples, described in Table I, are  GaAs double quantum well hole systems grown on (311)A substrates.     Each   wafer has a pair of  $150$ \AA~ GaAs  
  wells,   which are separated by AlAs barriers for $d\le300$ \AA, and a combination of AlAs and AlGaAs for $d>300$ \AA.  All the wafers are designed to have negligible interlayer tunneling.   
  Re$(\sigma_{xx})$ vs $B$ traces, like those in Fig. 1b, show  a dip  around  $\nu=1$   for the lower $\tilde{d}$ samples.  The absence of this 
 dip for $\tilde{d}\ge 1.69$ nearly    agrees with  earlier results \cite{eisnu1,spielman},  in which the $\nu=1$ state is present  for $\tilde{d}\le 1.8$.

\begin{table}[b]
\begin{center}  
\begin{tabular}{|c|c|c|c|c|}
\hline
Wafer   & d (\AA)& $p$ ($10^{10}{\rm cm}^{-2}$) & $\tilde{d}$& $\nu=1$ QHE\\
\hline
$M440$ & 225 & 3.00 & 1.38 & yes \\
\hline
$M465$ & 230 & 3.65 & 1.56 & yes \\
\hline
$M417$ & 260 & 2.70  & 1.51  & yes \\
\             &    \     & 3.05 &1.61 & \  yes  \\
\hline
$M433$ & 300 & 2.52  & 1.69  & no \\
\             &    \     & 2.85  & 1.80 &   no  \\
\hline
$M436$ & 450 & 2.40 & 2.47 & no \\
\hline
$M443$ & 650 & 2.85 & 3.89 & no \\
\hline
$M453$ & 2170 & 5.3 & 18 & no \\
\hline
\end{tabular}
 \end{center}
 \vspace*{-5pt} 
\caption{ Sample parameters in balanced states:  center-to-center well separation ($d$),   density per layer ($p$)  and effective separation $\tilde{d}=d/l_B$, where $l_B= (4\pi p)^{-1/2}$ is the magnetic length at total filling factor $\nu=1$.   The two $p$, $\tilde{d}$ for M417 and M433 are from different cooldowns.   The last column states if  a minimum in Re$(\sigma_{xx})$ vs magnetic field at $\nu=1$ is present. 
 \label{default} }
\end{table}

We performed microwave measurements using a method that has been described   earlier  
\cite{yewc,clibdep,clidensity}.  A   coplanar waveguide (CPW) transmission line,   on   top  of the sample, couples capacitively to the bilayer, as shown schematically in Fig. 1a. The CPW has a narrow, driven center conductor separated from grounded side planes by slots of width $W$. The  measurements proceed in a  high frequency, low loss limit of the CPW, in which the microwave field is only slightly perturbed by the conductivity of the bilayer. In this case  the in-plane microwave electric field  1) is mainly confined to the region immediately under the slots, and 2) is  essentially the same in both layers. We calculated the real part of diagonal conductivity from the transmitted power $P$ using a formula valid in this limit, when reflections are minimal, Re$(\sigma_{xx})=-W|\ln(P/P_0)|/2Z_0L$, where $P_0$ is the transmitted power with all carriers depleted from the bilayer, $Z_0=50\ \Omega$ is the characteristic impedance calculated for the  CPW geometry  with $\sigma_{xx}=0$, and $L$ is the length of the CPW.  The data, all taken around 60 mK,  were obtained in the low   power limit, in which further power reduction   did not affect the measured  Re$(\sigma_{xx})$.    A  voltage  applied between a  backgate and contacts to both layers was used  to control the layer densities.

The microwave measuring method precludes the use of front (top) gates to allow independent tuning  of the top and bottom layer densities.  
A metal film too close to the slots would effectively shield the bilayer from the required in-plane  microwave field. Hence, a balanced state is produced only by tuning  the backgate voltage  and can for each cooldown be produced at only one density.   For comparison with balanced bilayer states with layer densities $(p,p)$, we realize a single  layer state $(p,0)$ {\em in situ}, by reducing the total density, as evaluated from   QH    features,  by $50\%$ from that in the balanced state. 

Figure 1b shows    Re$(\sigma_{xx})$ vs $B$ for M465 and M433, in their balanced states, measured at $f=200$ MHz.   It shows that    
 the  main features observed \cite{tutucnu1,tutucnu1imbal,tutuchyst} in dc   transport are still readily observable with the transmission line.   M465, with $\tilde{d}\approx 1.56$ exhibits the $\nu=1$ IQHE, while M433, with $\tilde{d}\approx 1.8$, does not.     We used rf traces  
 like these to assess the total density and also to find the balanced states, by adjusting the backgate voltage,    to minimize the strength of the $\nu=1$  feature (for M440, M465, and M417) \cite{tutucnu1imbal}, or to minimize the hysteresis in  $B$   of QH states (for M433, M436, M443, and M453) \cite{tutuchyst}.

\begin{figure}
\includegraphics[width=3.2in]{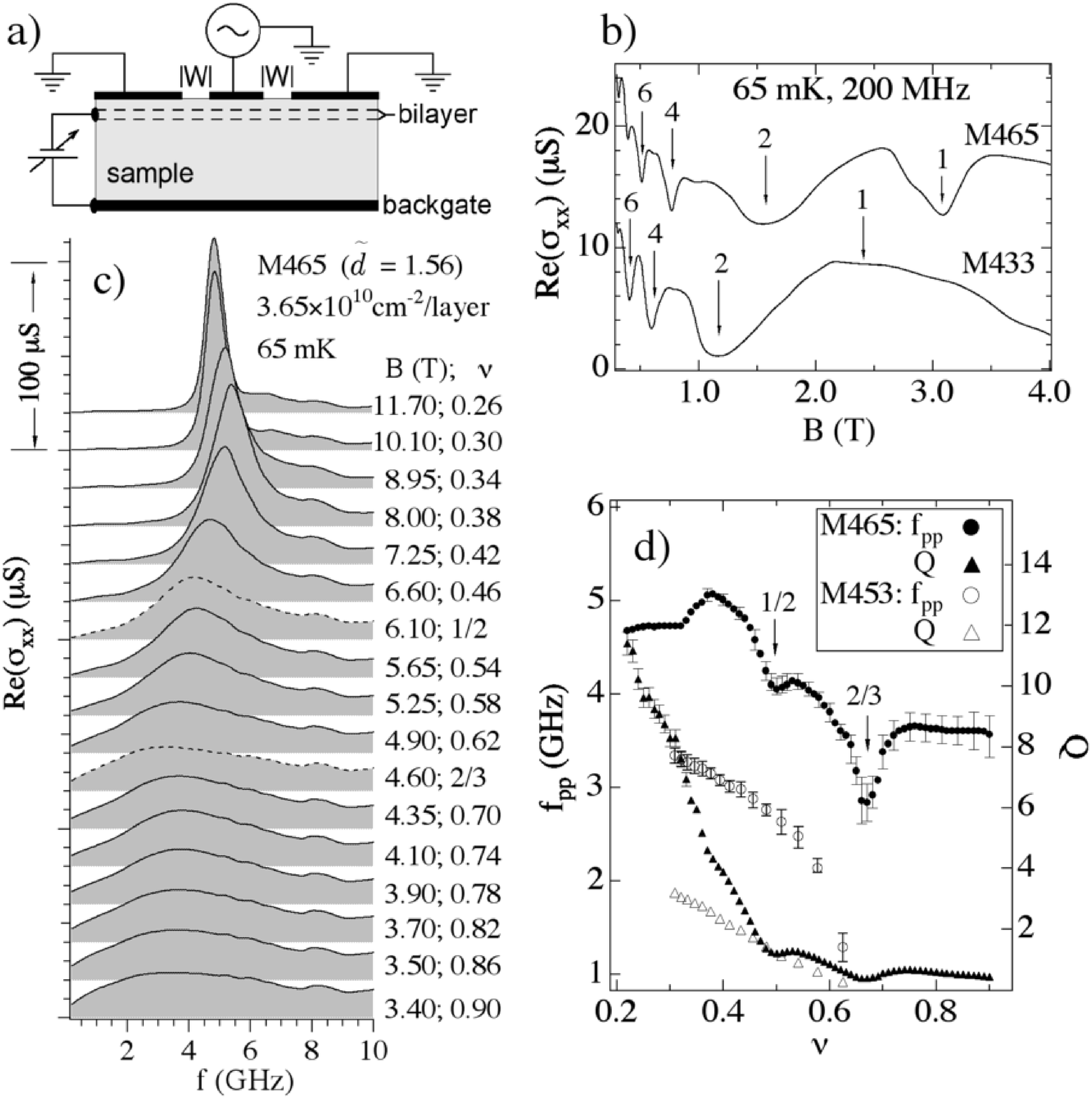}
\caption{\label{} a) Magnified cross section, not to scale, through the  transmission line and sample. b) Real part of diagonal conductivity  Re$(\sigma_{xx})$ vs. magnetic field, $B$, for  M465,   and M433, in their balanced states.   Total Landau  filling factors, $\nu$, are marked.   c)   Re$(\sigma_{xx})$ vs  frequency, $f$, for M465   in its balanced state, for many  $B$, (or   $\nu$).   Spectra are vertically displaced  proportional to $\nu$ and shaded downward to their zeros.  d)  peak frequency $f_{pp}$ and $Q$ vs $\nu$ ($Q$ is $f_{pp}$ divided by  full width at half maximum).     }\end{figure}

Figure 1c  illustrates  the evolution with $B$ (and $\nu$) of the spectra of 
M465 in its balanced state.  
Parameters from the spectra,  $f_{pp}$ and  $Q$ ($f_{pp}$ divided by full width at half maximum linewidth), are plotted  vs $\nu$  in Fig. 1d.     In M465,  and the other  two samples (M417 and M440) with the smallest $\tilde{d}$,  the  resonance is present for $\nu$ just below  the $\nu=1$ QH minimum, and sharpens dramatically as   $\nu$  decreases below about 0.5,    so that the  $Q$ vs $\nu$  curve  in Fig. 1d shows  a change of slope at that filling.   

In the samples with $\tilde{d}\ge1.69$, the resonance appears as $\nu$ decreases below about 0.6, with gradually increasing $f_{pp}$ and $Q$.    
This resonance development  is typical of that seen at the same per-layer filling, previously \cite{clibdep} in low density p type single layers,  and  in the  present  samples in their single layer $(p,0)$ states.    \  $f_{pp}$ and $Q$ for M453, which has the largest $\tilde{d}$, and therefore essentially independent layers, are
 plotted vs $\nu$ in Fig.~1d as well. 

 M465 shows   clear    dips in $f_{pp}$ vs $\nu$ around $\nu= 1/2 $ and $2/3$  in the plot in Fig.~1c .   (M417 and M440  have  weak  dips in $f_{pp}$ at  2/3 or 1/2,  but the dips   in those samples are not nearly as clear as   in M465, possibly  due to the higher  $p$ of M465.)  
The resonance is well-developed at these fillings, which  are well  within the  insulating phase of this sample,  so we do not interpret the dips  as FQHE liquid ground states  \cite{eis12,suen12,suen1232,hamilton32}, but rather as evidence that some of the  correlations related to the FQHE are present in the pinned BWC.    In general, as we discuss in more detail below,
 as  Wigner solid moduli increase, the resonance frequency decreases   \cite{clidensity,chitra,fertig,fogler};  hence the dips in $f_{pp}$  
 correspond either to { \em increases} in    BWC stiffness, or to decreases in the effective carrier-disorder interaction.    A decrease in  resonance frequency 
 as  $\nu $ approaches an FQHE  has been observed \cite{yongab} in single layer samples.   
The 1/2 FQHE has no analog in the single layer case, and is  interlayer correlated, though the 2/3 feature can   be explained  with intralayer correlation.   Composite fermion 
 theories  \cite{narevichfertig,jaincfwc}    have predicted that   correlations  related to the FQHE are important in single layer, low $\nu$ Wigner solids.

\begin{figure}[t]
\includegraphics[width=3.2in]{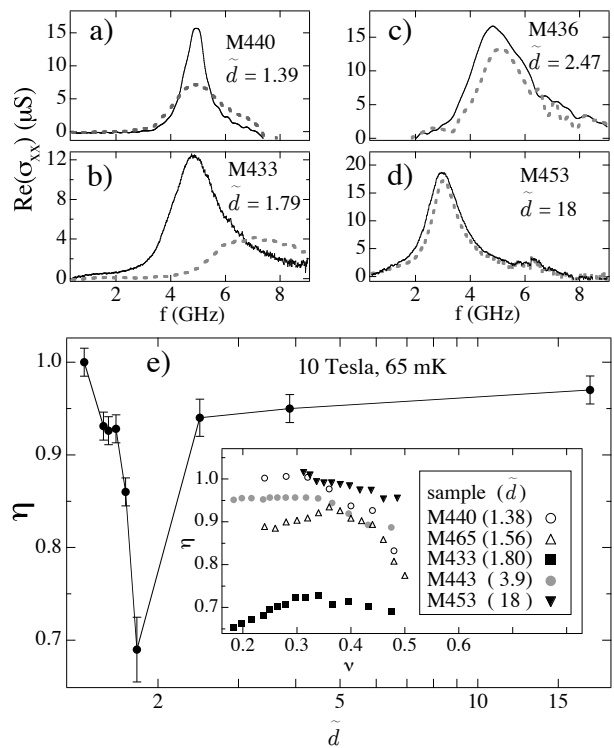}
 \vspace*{-5pt}
\caption{\label{} a)- d) For four samples,the  spectrum in the balanced state with layer densities  $(p,p)$ (solid line)  is plotted  along with the spectrum (with   the measured Re$(\sigma_{xx})$ multiplied by two) of the single layer state $(p,0)$ (dotted line),  all at magnetic field $B=10$ T.   (e)   $\eta= f_{pp}/f_{p0}$ at $10$ T  vs $\tilde{d}= d (4\pi p)^{1/2} $, where $f_{pp}$, $f_{p0} $  are the peak frequencies for the   $(p,p)$ and    $(p,0)$ states and  $d$ is layer separation. Inset: $\eta$ vs $\nu$, the  total filling in the bilayer $(p,p)$ state,  for five samples.}
\end{figure} 
The main results of this paper follow from direct comparison, for each sample,  of the balanced $(p,p)$ and single layer $(p,0)$ states.
  Spectra  from these pairs of states
are shown for four samples, at $B=10$ T in Fig.  2a-d. The bilayer spectra are shown as solid lines, and the single layer spectra are dashed.     \resxx\ from the single layer states is doubled to facilitate comparison.   The M453 spectra   in Fig. 2d  are nearly identical, as expected for independent layers, and  not surprising considering the large  $\tilde{d}\approx 18$ for that sample.   
Important for the  interpretation below, the nearly identical spectra also indicate that  
the disorder statistics  relevant to the   pinning mode are essentially the same  in the top and bottom wells.   The disorders of  the two layers should likewise be similar for  all the samples, since  they all  had similar growth characteristics,  such as  asymmetrical doping and interfacial compositions.

Hence we  interpret the differences between the $(p,p)$ and $(p,0)$ spectra in Fig. 2a-d as due to changes in  interlayer interaction and correlation.
Relative to the $(p,0)$ spectra,  the   $(p,p)$ spectra shift slightly  to lower frequency  as $\tilde{d}$ decreases  
down to  2.47, as shown in Fig.~2c.  At  $\tilde{d} \approx 1.8$ though, for M433 as shown in Fig. 2b,  the $(p,p)$ spectrum is markedly shifted downward in frequency, and is  stronger and sharper. But decreasing  $\tilde{d}$ further (even through a  different cooldown of the same  M433 sample)  reduces the downward shift of   $f_{pp}$ relative to $f_{p0}$, though the $(p,p)$ spectra remain much sharper than the $(p,0)$ spectra, as seen in Fig.~2a. To summarize, Fig.~2e shows  the  ratio $\eta=f_{pp}/f_{p0}$ vs $\tilde{d}$; this curve has  a striking minimum at $\tilde{d}\approx 1.8$.  Inset in Fig. 2e is a graph of $\eta$ vs $\nu$, the total filling of the $(p,p)$ states, for several samples.  This graph shows that the minimum in $\eta$ vs $\tilde{d}$ remains, whether $B$ or $\nu$ is fixed, as long as $\nu\lesssim 0.5$.   For $\tilde{d}\le1.8$, as shown for  M465 in Fig. 1b and c,
 resonances  become markedly sharper below  $\nu\approx 0.5$.

We interpret the $\eta$ vs $\tilde{d}$ curve 
as a result of two competing effects.    The first effect is driven by carrier-carrier interaction and must be present in any weakly pinned Wigner crystal, bilayer or single layer.  In this effect, when carrier-carrier interaction 
 is increased, by  decreasing their spacing (or increasing their overall density),     the   resonance frequency decreases.         In single layers at fixed $B$,  at which  the resonance is well developed,  as carrier density $n_s$ is decreased,   the peak frequency $f_{pk}$  always increases and  the resonance broadens.   Typically \cite{clidensity,yewc},
$f_{pk}\propto n_s^{-\gamma}$, with $\gamma\approx 3/2$ for higher $n_s$ giving way to $\gamma\approx 1/2$ at lower $n_s$; the present samples in  single layer states      
all have $\gamma\approx 1/2 \pm 10\%$.  The interpretation in weak pinning  \cite{clidensity,chitra,fertig,fogler} of  the increase of $f_{pk}$  with decreasing $n_s$ is that reduction   of carrier-carrier interaction ({\em i.e.}, the crystal stiffness)  causes the carriers to adjust to positions which essentially fall more deeply into the impurity potential.  This increases the pinning energy per carrier, and the restoring force on the carriers, hence $f_{pk}$.

 We interpret the  decrease of $\eta$ with decreasing  $\tilde{d}$,  for  $\tilde{d}\ge 1.8$,  as due to this carrier-carrier interaction effect, within an AFMBWC.  The result of this effect, on going from $(p,0)$ to $(p,p)$ in the  limit of small $\tilde{d}$, is analogous to doubling the 
 areal density of a single layer, and for $\gamma\sim 1/2$ gives 
 $\eta=2^{-\gamma}\approx 0.71$.   This    agrees with the $\eta$ we measure for  $\tilde{d}\approx 1.8$.            
 The sharp increase in $\eta$ as $\tilde{d}$ goes below 1.8 is  not readily explainable in terms of the carrier-carrier interaction effect.   
     Transitions between different types of  AFMBWC  are predicted by the theories \cite{ho,zhengfertig,peeters}   
and if  $ \tilde{d}$ is near a transition, the BWC can conceivably soften (multiple low energy arrangements become possible),  producing  some increase of $\eta $ around particular $\tilde{d}$.   
It is not likely though,  that even around a 
transition between AFMBWC phases,  $\eta$ would be as
close to unity as it is  in Fig. 2e at the smallest $\tilde{d}$ values.

The  second  competing effect that we use to explain  $\eta$ vs $\tilde{d}$    is driven by interlayer correlation, is present only in the FMBWC, and was  considered theoretically by Chen \cite{yong}.   Chen found that  in an FMBWC  pinning is enhanced  when   there is disorder that is  spatially correlated in the planes of the top and bottom layers.  At sites where impurities or interfacial features induce   local interlayer tunneling \cite{tutucnu1}, such spatial correlation would naturally result.     When  this disorder enhancement is  considered \cite{yong} along with the competing carrier-carrier interaction effect, $\eta$ as large as $2^{1-\gamma}$ is possible, so the  transition to an FMBWC is sufficient to explain the  
  increase of $\eta$  with decreasing $\tilde{d}$ seen in Fig. 2e for  $\tilde{d}\lesssim 1.8$.        Even within the FMBWC,   an increase of $\eta$ as $\tilde{d}$ decreases is expected, since the smaller $\tilde{d}$ would increase the interlayer-spatially correlated component of effective disorder.

The data then indicate that  $\tilde{d}^*$,  the critical $\tilde{d}$  below which the FMBWC is present, is  around  $1.8$.
Theories  with small but finite tunneling 
 predict   smaller   $\tilde{d}^*$,   around 0.4 \cite{ho,zhengfertig}.    A possible explanation of the discrepancy lies in the increased pinning experienced by the FMBWC, since the  pinning energy can  stabilize the FMBWC against the more weakly pinned AFMBWC phases that succeed it at larger $\tilde{d}$.   It is likely that pinning energy   plays a role in stabilizing  single layer Wigner crystals     against the   FQHE liquid at high fillings \cite{pzl},   and against melting at elevated temperatures \cite{yongmelt}.
A possible  clue to the presently estimated  $\tilde{d}^*\approx  1.8$    is that it is close to  the maximal  $\tilde{d}$ values below which  the interlayer correlated QH states at $\nu=1$ and $1/2$   exist,    respectively   $\tilde{d}\approx 1.8 $  \cite{eisnu1,spielman} and $2$ \cite{eis12}.

 In sum, our systematic studies of pinning modes of BWC have found effects of  interlayer interaction and give evidence for  BWC  interlayer correlations in which  the  wave function of a carrier spreads evenly between the two layers.  Correlations related as precursors to the bilayer 1/2  and 2/3  FQHEs are also present in the  BWC. 
 
We thank N. Bonesteel, H. A. Fertig  and Kun Yang for discussions, and acknowledge 
the  support of AFOSR and  DOE.   NHMFL is supported by NSF grant DMR-0084173,  the State of Florida, and  DOE.

\end{document}